\documentclass[a4paper, 10 pt, conference]{cras} 
\usepackage[utf8]{inputenc}
\IEEEoverridecommandlockouts                       
\usepackage{mathtools}
\usepackage{geometry}
\usepackage{newtxmath,newtxtext}
\usepackage{authblk}
\usepackage{pgfplots}
\usepackage{subfig}
\usepackage{graphicx}
\pgfplotsset{compat=newest} 
\pgfplotsset{plot coordinates/math parser=false} 
 
\title{\Large \bf
Scribble-Based Interactive Segmentation of Medical Hyperspectral Images
\thanks{%
This project has received funding from NIHR [NIHR202114], Wellcome/EPSRC [WT203148/Z/16/Z; NS/A000049/1], European Union's Horizon 2020 [101016985].
For the purpose of open access, the authors have applied a CC BY public copyright license to any Author Accepted Manuscript version arising from this submission.
TV and JS are co-founders and shareholders of Hypervision Surgical.
}} 
\author[1]{\large Zhonghao Wang}
\author[1]{\large Junwen Wang} 
\author[1]{\large Charlie Budd} 
\author[1,2]{\large Oscar MacCormac}
\author[1,2]{\\\large Jonathan Shapey}
\author[1]{\large Tom Vercauteren} 
\affil[1]{\small\textit{School of Biomedical Engineering \& Imaging Sciences, King’s College London, UK}}
\affil[2]{\small\textit{Department of Neurosurgery, King's College Hospital, London,  UK}}
\begin{document}

\maketitle

\begin{figure*}[!t]
    \centering
    \subfloat[Hyperspectral image]{\includegraphics[width=0.195\textwidth]{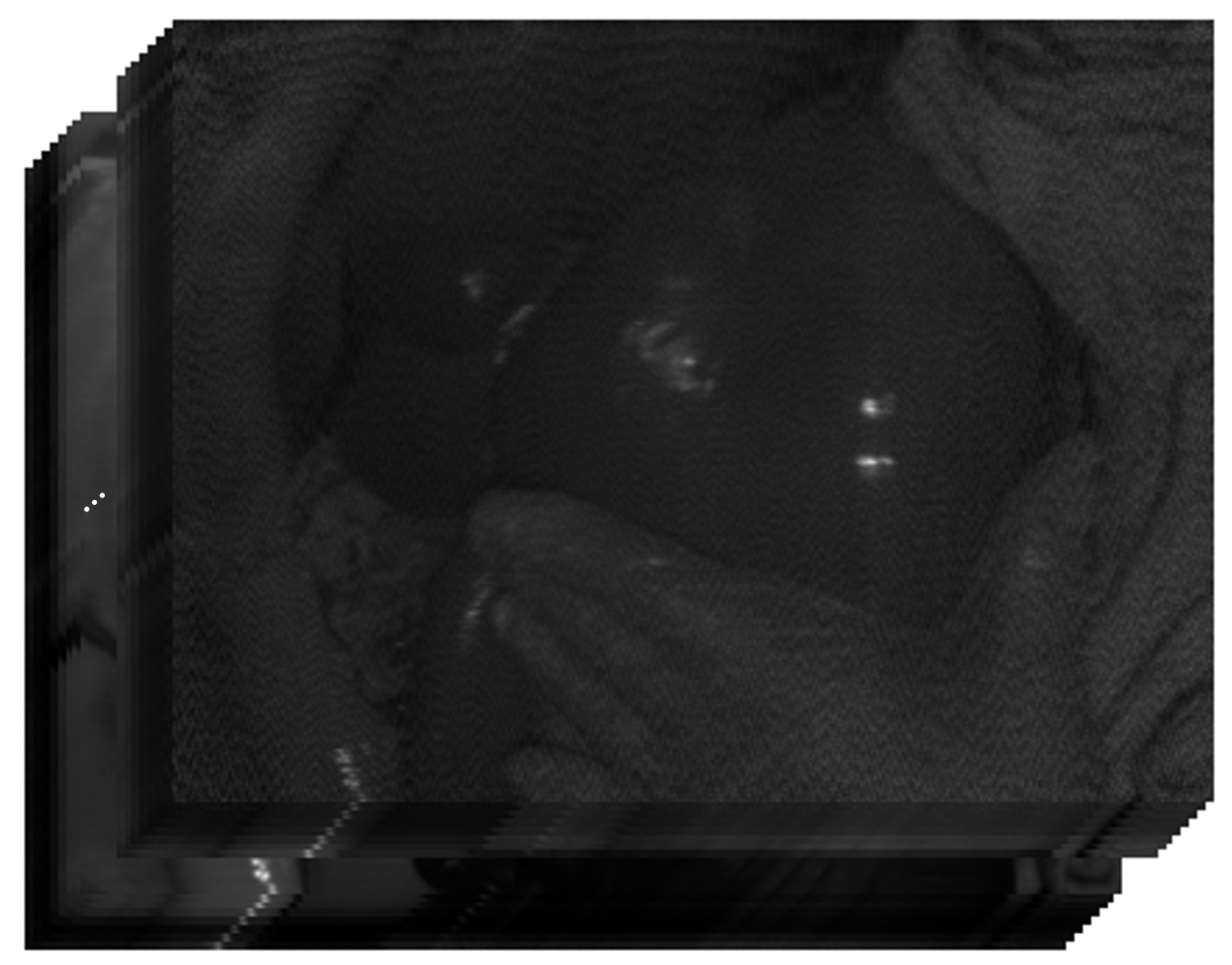}\label{fig:image1}}
    \hfill
    \subfloat[Extracted features]{\includegraphics[width=0.195\textwidth]{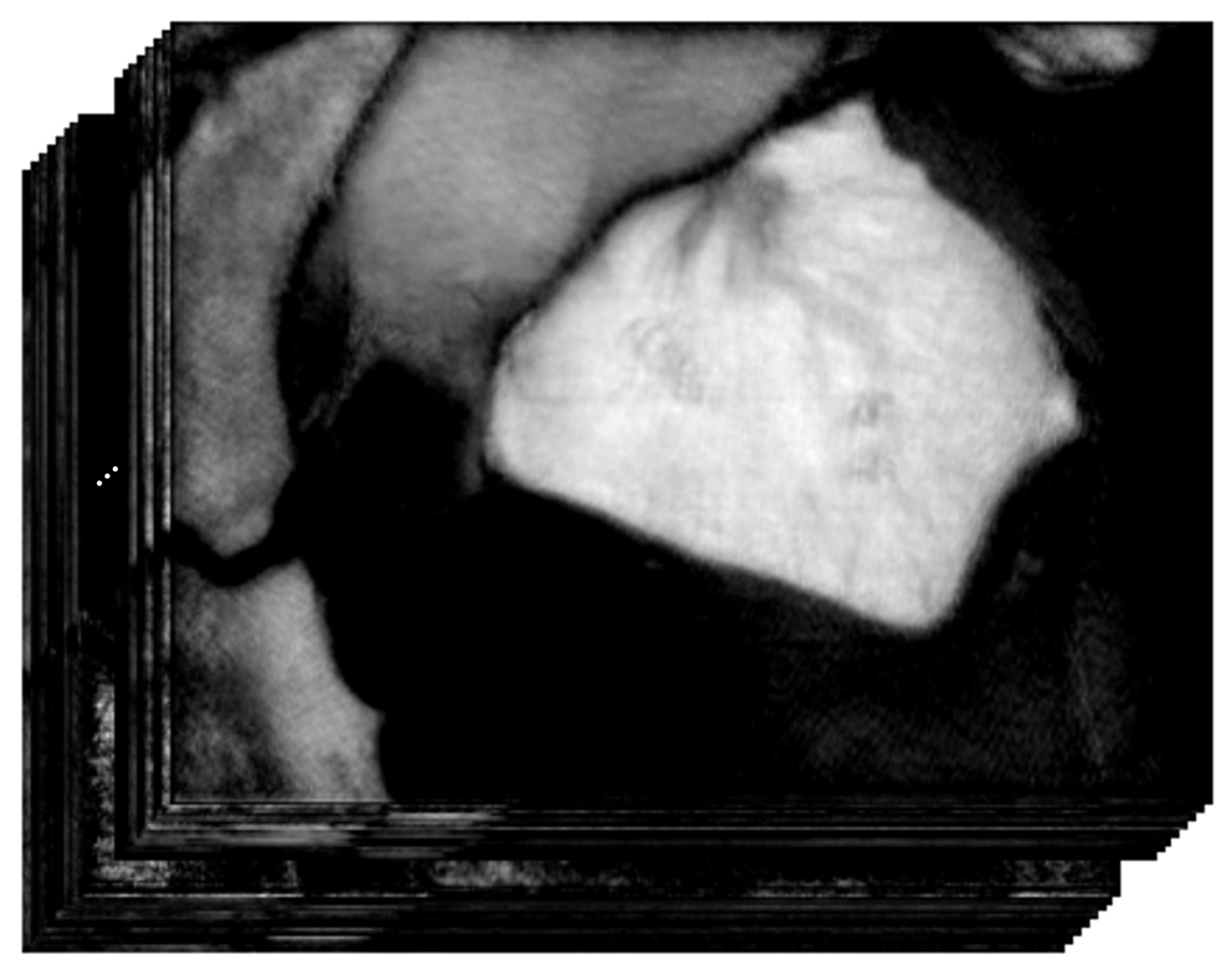}\label{fig:image2}}
    \hfill
    \subfloat[Scribbles]{\includegraphics[width=0.2\textwidth]{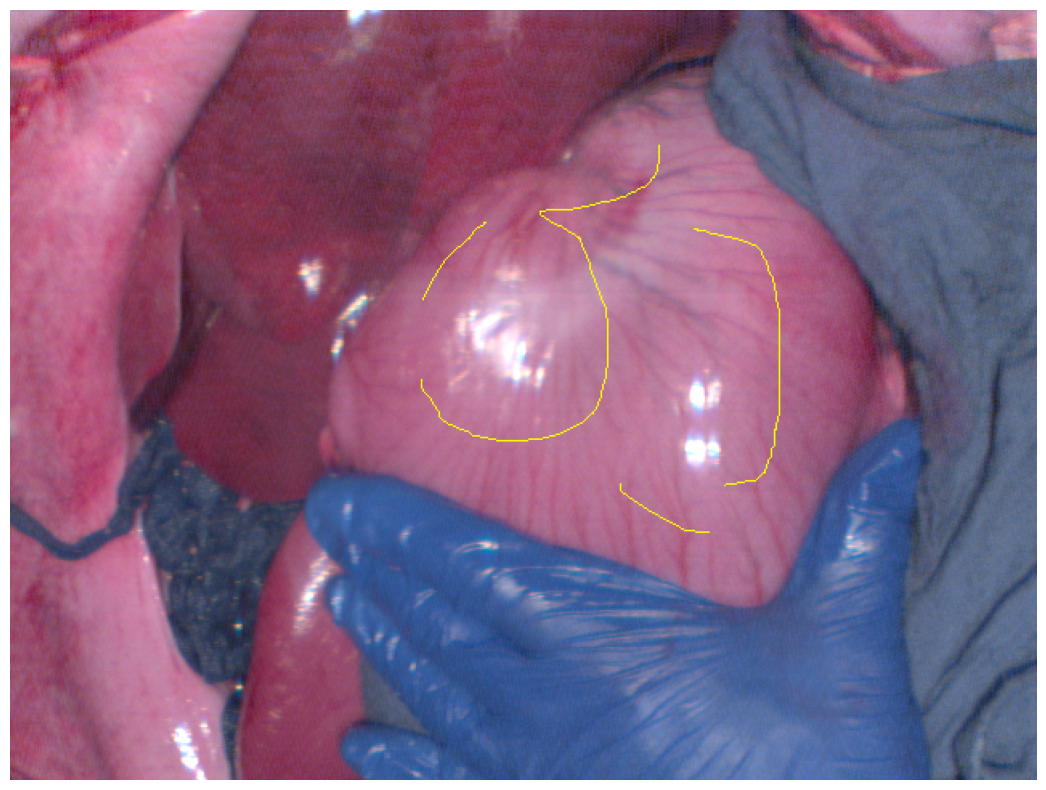}\label{fig:image3}}
    \hfill
    \subfloat[Geodesic distance map]{\includegraphics[width=0.2\textwidth]{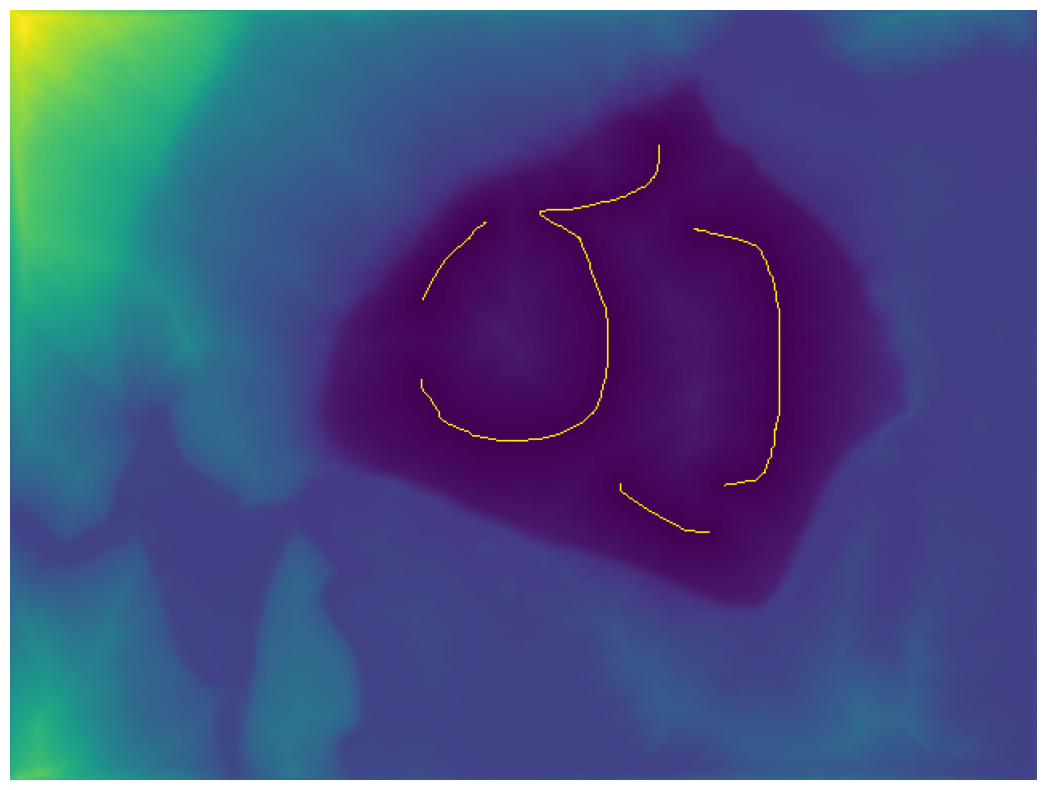}\label{fig:image4}}
    \hfill
    \subfloat[Segmentation result (blue) and ground truth (white)]{\includegraphics[width=0.2\textwidth]{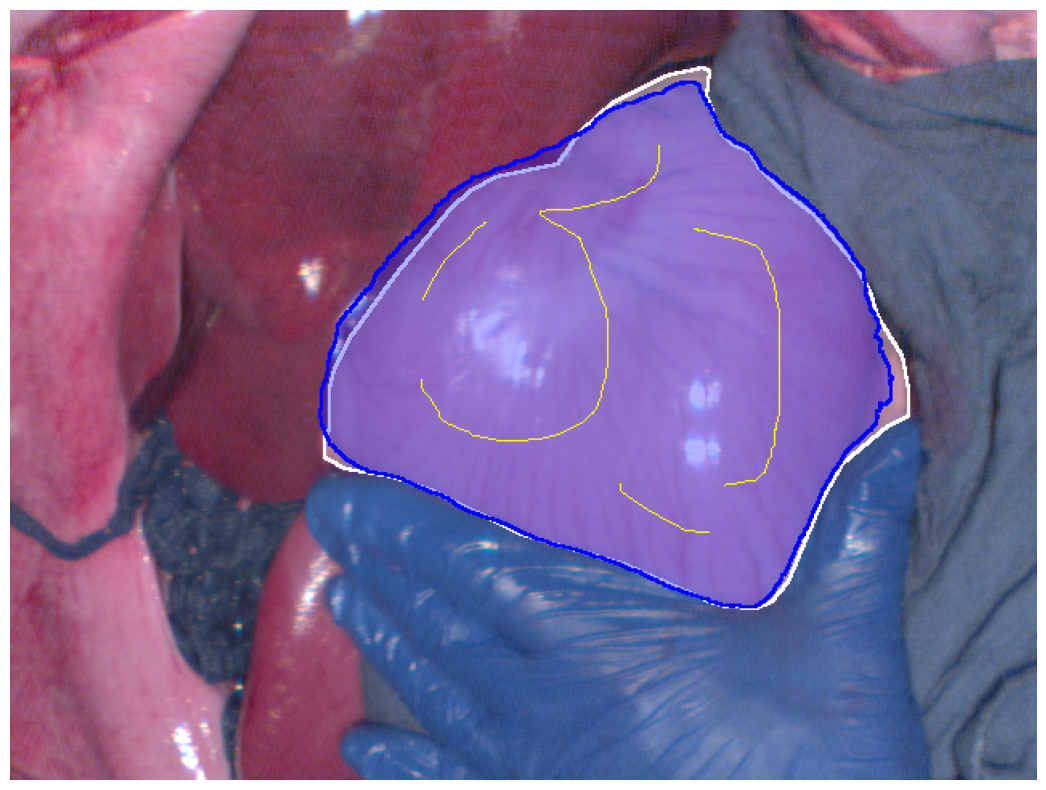}\label{fig:image5}}
    \caption{Overview of the scribble-based hyperspectral image interactive segmentation process.}
    \label{fig:five_images}
\end{figure*}
\thispagestyle{empty}
\pagestyle{empty}

\section*{INTRODUCTION}
Hyperspectral imaging (HSI) is an advanced medical imaging modality that captures optical data across a broad spectral range, providing novel insights into the biochemical composition of tissues.
HSI may enable precise differentiation between various tissue types and pathologies, making it particularly valuable for tumour detection, tissue classification, and disease diagnosis \cite{clancy2020surgical}.

Deep learning-based segmentation methods have shown considerable advancements, offering automated and accurate results. However, these methods face challenges with HSI datasets due to limited annotated data and discrepancies from hardware and acquisition techniques~\cite{clancy2020surgical,studier2023heiporspectral}.
Variability in clinical protocols also leads to different definitions of structure boundaries. Interactive segmentation methods, utilizing user knowledge and clinical insights, can overcome these issues and achieve precise segmentation results \cite{zhao2013overview}.

This work introduces a scribble-based interactive segmentation framework for medical hyperspectral images. The proposed method utilizes deep learning for feature extraction and a geodesic distance map generated from user-provided scribbles to obtain the segmentation results. The experiment results show that utilising the geodesic distance maps based on deep learning-extracted features achieved better segmentation results than geodesic distance maps directly generated from hyperspectral images, reconstructed RGB images, or Euclidean distance maps.

\section*{MATERIALS AND METHODS}
\subsubsection*{Preprocessing and deep learning-based feature extraction}
As shown in Fig.~\ref{fig:image1}, hyperspectral images consist of one spectral and two spatial dimensions. Before inputting the hyperspectral image into the neural network, L1 normalization along the spectral dimension is applied to reduce the influence of illumination variations. A U-Net-based deep learning network is then used for feature extraction from hyperspectral images~\cite{ronneberger2015u}. The U-Net model follows an encoder-decoder structure, where the encoder progressively reduces the spatial dimensions of the input, capturing higher-level features, and the decoder progressively restores the spatial dimensions while combining them with the corresponding features from the encoder. The final output from the decoder is used as the extracted features (Fig. \ref{fig:image2}). This approach ensures that the features retain both high-level semantic information and fine-grained spatial details, which are crucial for accurate segmentation.

\subsubsection*{Geodesic distance map generation}
The geodesic distance map is generated based on user-provided scribbles and the extracted features~\cite{asad2022fastgeodis}.
Let $S$ be the set of pixels in the user-provided scribbles and $i$ be a pixel in an image $\mathbf{I}$.
The unsigned geodesic distance from $i$ to $S$ is:
\[
G(i, S, \mathbf{I}) = \min_{j \in S} D_{\text{geo}}(i, j, \mathbf{I})
\] where $D_{\text{geo}}(i, j, \mathbf{I}) = \min_{p \in \mathcal{P}_{i,j}} \int_{0}^{1} \| \nabla \mathbf{I}(p(s)) \cdot \mathbf{u}(s) \| ds$;
$\mathcal{P}_{i,j}$ is the set of all paths between pixel $i$ and pixel $j$;
$p$ is one feasible path parameterized by $s \in [0,1]$;
and the vector $\mathbf{u}(s) = \frac{p'(s)}{\|p'(s)\|}$ is a unit vector tangent to the
%direction of the 
path. 

%As shown in Fig.~\ref{fig:image3}, the user draws yellow scribbles as input 
Fig.~\ref{fig:image3} shows exemplar user scribbles drawn
on the reconstructed RGB image, while Fig.~\ref{fig:image4} shows the generated geodesic distance map.
The unsigned geodesic distance increases
%from dark blue to bright yellow.
%The geodesic distance can
away from the scribbles while
capturing the spatial relationships and inherent contrast within the feature.
This helps better differentiate neighbouring pixels with different appearances and improve label consistency in homogeneous regions \cite{criminisi2008geos,Wang:PAMI:2019}.

\subsubsection*{Extending scribbles to segmentation result}
After generating the geodesic distance map, it is normalized to a range of 0 to 1. A binary classification is generated according to the user-controlled threshold, where pixels with geodesic distance values below the threshold are labelled as part of the region of interest. The user can adjust the threshold value to extend the scribbles into an appropriate segmentation result.
%As shown in
In Fig.~\ref{fig:image5}, the white region represents the ground truth (indicating a pig's stomach), while the blue region shows the segmentation result when the threshold is adjusted to achieve the maximum Dice coefficient.
The user can also lower the threshold to obtain a more conservative segmentation result.
This method is very flexible and also reduces the user's annotation workload.

\section*{RESULTS}
The HeiPorSPECTRAL dataset \cite{studier2023heiporspectral} was utilised for the experiments. Initially, the U-Net model was trained using data from 10 subjects, excluding subject P086. After training, the scribble-based interactive segmentation framework was evaluated on the untrained data from subject P086. Comparisons were made between the geodesic distance maps generated from the extracted features, the original hyperspectral image, the reconstructed RGB image, and the Euclidean distance map.
These comparisons aimed to evaluate the accuracy of the proposed method using the extracted features against traditional approaches, highlighting the benefits of integrating deep learning-based feature extraction with geodesic distance mapping.
\begin{figure}[!t]
    \centering
    \begin{minipage}{\linewidth}
        \centering
        \begin{minipage}{0.49\linewidth}
            \centering
            \subfloat[Deep learning-based feature best Dice segmentation result]{\includegraphics[width=\linewidth]{pics/max_dice_fea.png}\label{fig:subfig1}}
        \end{minipage}
        \hfill
        \begin{minipage}{0.49\linewidth}
            \centering
            \subfloat[Hyperspectral image best Dice segmentation result]{\includegraphics[width=\linewidth]{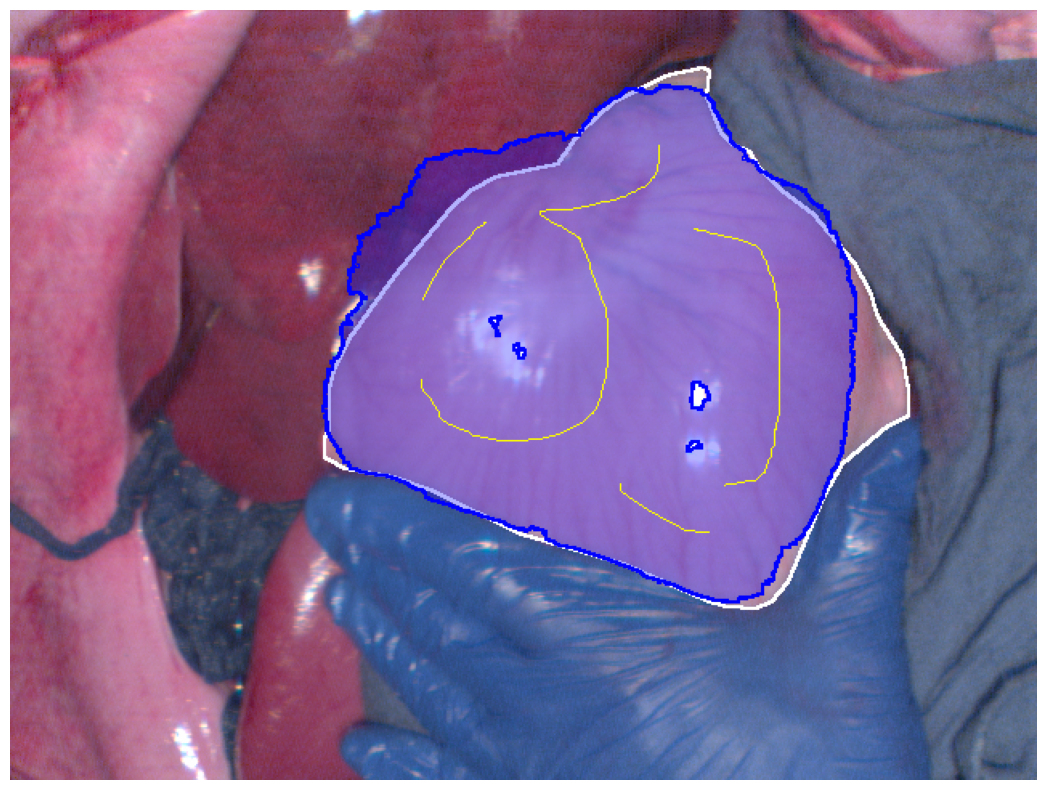}\label{fig:subfig2}}
        \end{minipage}
    \end{minipage}
    \vfill
    \begin{minipage}{\linewidth}
        \centering
        \begin{minipage}{0.49\linewidth}
            \centering
            \subfloat[Reconstructed RGB best Dice segmentation result]{\includegraphics[width=\linewidth]{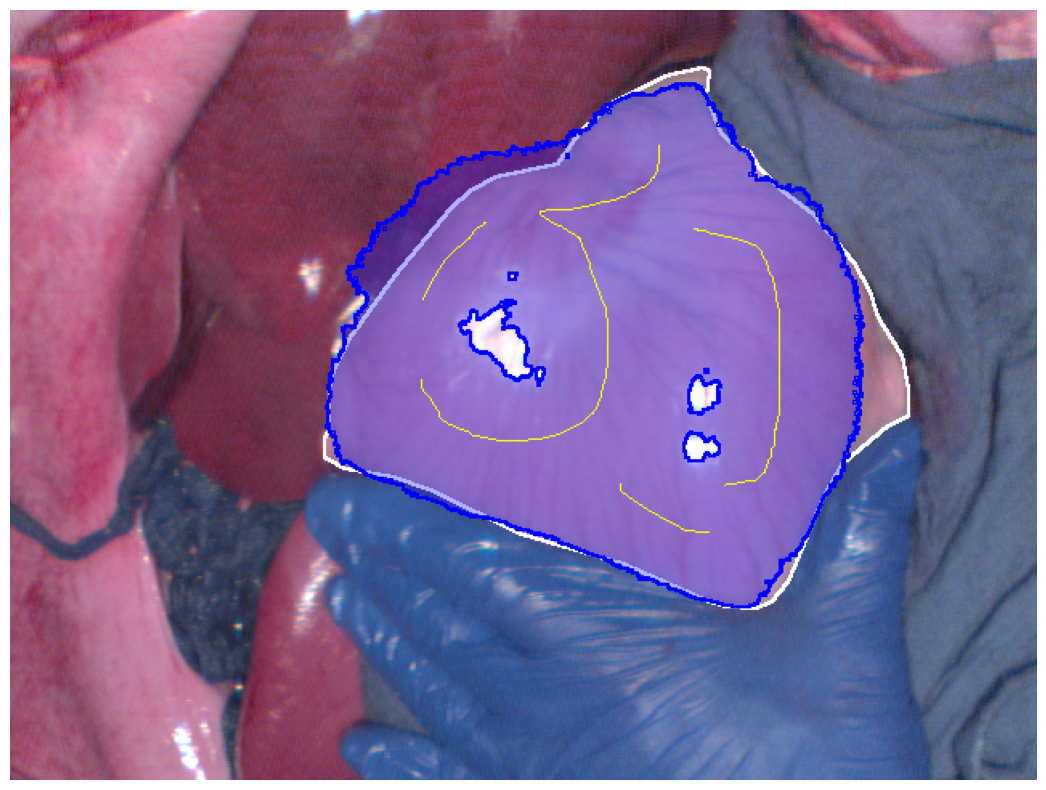}\label{fig:subfig3}}
        \end{minipage}
        \hfill
        \begin{minipage}{0.49\linewidth}
            \centering
            \subfloat[Euclidean distance map best Dice segmentation result]{\includegraphics[width=\linewidth]{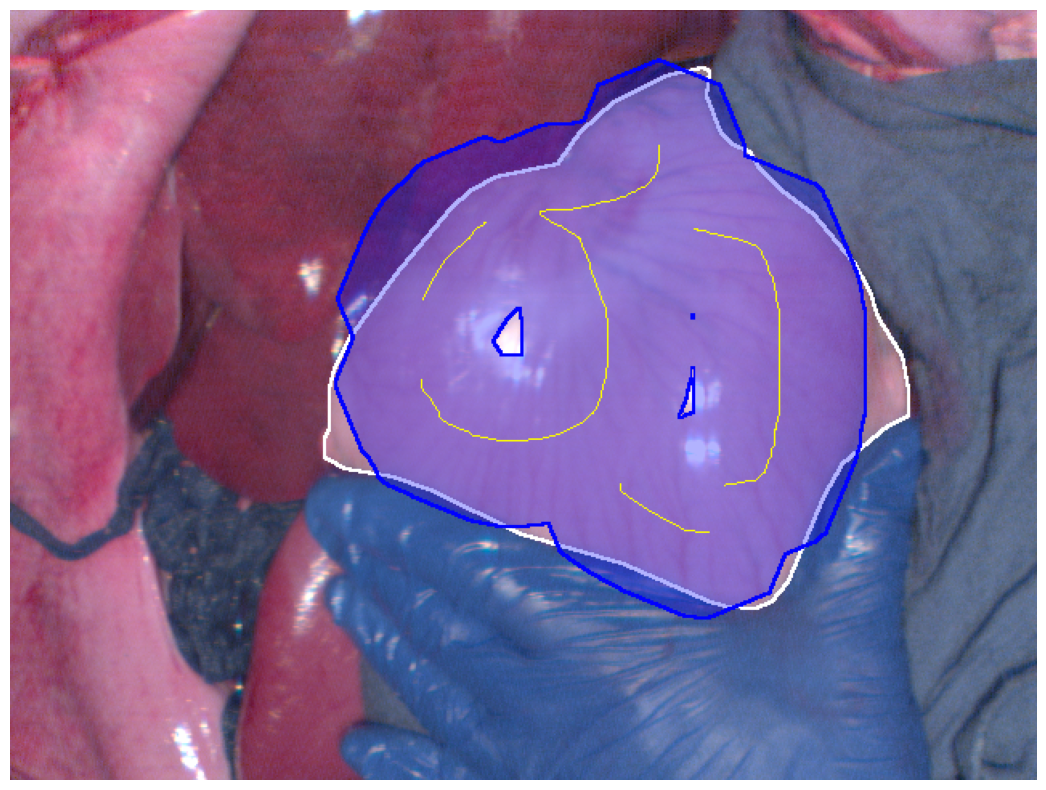}\label{fig:subfig4}}
        \end{minipage}
    \end{minipage}
    \vfill
    \subfloat[Dice score variation with threshold adjustment in different methods]{\includegraphics[width={\linewidth}]{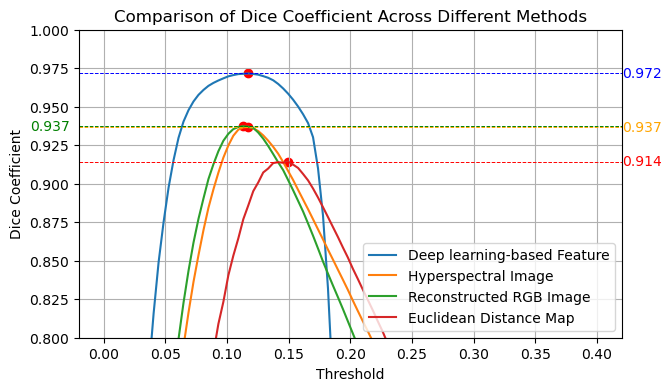}\label{fig:subfig5}}
    \caption{Segmentation results (blue regions) at the best Dice coefficient for different methods and ground truth (white regions), along with the curves of Dice coefficient variation with threshold adjustments.}
    \label{fig:combined}
\end{figure}
In Fig.~\ref{fig:combined}, the segmentation results at the best Dice for the four different methods are displayed, along with the variation in Dice coefficients with threshold adjustments.
From Fig.~\ref{fig:subfig5}, it can be seen that the Euclidean distance method has the lowest maximum Dice score of 0.914.
The methods using geodesic distance maps generated from hyperspectral images and reconstructed RGB images achieve similar maximum Dice values, both higher than the Euclidean method. The deep learning-based feature geodesic distance map method achieves the highest maximum dice coefficient.

For the 575 images in P086, automated scribbles were generated by skeletonizing the annotation results, and segmentation results were produced using pre-mentioned four different methods. Among these, the deep learning-based feature geodesic distance map method achieved the highest average max Dice score of 0.842, demonstrating its superior performance.

\section*{CONCLUSIONS AND DISCUSSION}
This work introduces a scribble-based interactive segmentation framework for medical hyperspectral images, taking advantage of deep learning for feature extraction and geodesic distance maps for accurate segmentation results. The method demonstrated superior performance compared to traditional methods, offering enhanced accuracy and flexibility, thereby reducing user workload. Future work will focus on further improving the framework and extending its application to other datasets.

%\nocite{*}
\bibliographystyle{IEEEtran}
\bibliography{CRAS}

\end{document}